\documentstyle[emulateapj,epsf]{article}

\def\be{\begin{eqnarray}}
\def\ee{\end{eqnarray}}

\def\etal{{\it et al.$\,\,$}}

\def\thvec{{ \mbox{\boldmath $\theta$} }}

\def\ds{{D_{\rm s}}}

\def\veffperp{{ V_{\rm eff, \perp}}}
\def\veffperpvec{{ \mbox{\boldmath $V_{\rm eff,\perp}$}}}
\def\veffperpvechat{{ \mbox{\boldmath $\hat V_{\rm eff,\perp}$}}}
\def\vismperpvec{{ \mbox{\boldmath ${V_{\rm screen}}_\perp$}}}
\def\vpperpvec{{ \mbox{\boldmath  ${V_{\rm p}}_\perp$}}}
\def\vobsperpvec{{ \mbox{\boldmath ${V_{\rm obs}}_\perp$}}}

\newcommand{\eg} {e.g.\ }

\begin{document}
\pagestyle{myheadings}



\title{Faint Scattering Around Pulsars:  Probing the Interstellar 
Medium on Solar System Size Scales}

\author{D. R. Stinebring\altaffilmark{1}, 
        M. A. McLaughlin\altaffilmark{2},
        J. M. Cordes\altaffilmark{2},
        K. M. Becker\altaffilmark{1},
        J. E. Espinoza Goodman\altaffilmark{1},
        M. A. Kramer\altaffilmark{1},
        J. L. Sheckard\altaffilmark{1},
        T. C. T. Smith\altaffilmark{1}}

\altaffiltext{1}{Department of Physics, Oberlin College, Oberlin, OH 
44074} 
\altaffiltext{2}{Dept. of Astronomy and Space Sciences Center, 
       Cornelll University, Ithaca, NY 14853}

\authoraddr{Address correspondence regarding this manuscript 
to: 
		Dan Stinebring
		Dept. of Physics
		Oberlin College
		Oberlin, OH 44074}

\begin{abstract}
We have made high-resolution, high-sensitivity dynamic spectra of a 
sample of strong 
pulsars at 430~MHz with the Arecibo radiotelescope.
For 4 pulsars we 
find faint but sharply delineated features in the secondary 
spectra.  These are examples of the previously observed ``crisscross'' or ``multiple 
drift slope'' phenomenon presumed to be due to multiple imaging of the 
pulsar by the interstellar medium.  The unprecedented resolution and 
dynamic range of our observations allow a deeper level of analysis.  
Distances to the dominant scattering 
screen along the line of sight are determined and are shown to agree 
well with those inferred from other scintillation phenomena.
Multiple imaging of the pulsar by the ISM is required.
A compact central image surrounded by a faint scattering halo, roughly 
circularly symmetric, is consistent with the data.
Scattering from filaments may also be consistent.
The angular extent of the scattering material parallel to the 
direction of the pulsar velocity is $\sim 5$~mas,  
corresponding to a linear extent of $\sim 2$~AU.
Further observations of these features should allow better 
discrimination between models and an identification of the scattering 
structures. 
\end{abstract}


\vspace{.25in}


\section{Introduction}

Scattering by density inhomogeneities in the ionized interstellar 
medium causes multiple rays to interfere, producing frequency 
structure in the spectra of spatially coherent sources such as pulsars 
\cite{sch68,ric90}.
The relatively high transverse 
velocities of most pulsars cause the spectra to change significantly on 
timescales of minutes to hours.
When the spectrum is monitored over time, the resulting two-dimensional array or 
``dynamic spectrum''  is often dominated by a random pattern of 
scintillation maxima with characteristic bandwidth and timescale 
\cite{cwb85}.
On occasion, however, more organized features are present. The best 
studied of these are periodic fringe patterns in the dynamic 
spectra due to the interference between two or more discrete bundles of rays 
or, equivalently, rays arriving from multiple images on the sky 
\cite[ hereafter RLG]{cw86,wc87,gbr99,rlg97}.

A less well-studied pattern in pulsar dynamic spectra
consists of a crisscross of intensity maxima that
are variably spaced perpendicular 
to the ridge lines of the maxima.  
Cordes and Wolszczan (1986) drew attention to these ``multiple drift 
rates'' in dynamic spectra, building on earlier work
(Hewish 1980; Roberts \& Ables 1982; Hewish, Wolszczan, \& Graham 
1985; Cordes, Weisberg, \& Boriakoff 1985). 
\nocite{hew80}
\nocite{ra82}
\nocite{hwg85}
\nocite{cwb85}
Numerous examples of the phenomenon are evident in the dynamic spectra
presented by Gupta, Rickett, \& Lyne (1994) in their multi-year study 
of 8 pulsars,
\nocite{grl94}
but the autocorrelation analysis that they employed is not well 
suited to its full elucidation.
The remarkable episode of ``fringing'' in the dynamic spectra
of PSR~B0834+06 that RLG explore is also accompanied at all
epochs by this phenomenon.

The central result of our study is to show that this 
crisscross pattern has a much clearer representation in transform 
space, where it is seen to be a high-Q phenomenon with a 
characteristic parabolic curvature.
We discuss the constraints that the presence of this pattern places on the
image geometry and
speculate on the physical structures that might give rise to this 
scattering geometry.

\section{Observations and Data Analysis}
Our observations were made at the Arecibo Observatory\footnote{The 
Arecibo Observatory is part of the National Astronomy 
and Ionosphere Center, which is operated by Cornell University under 
a cooperative agreement with the National Science Foundation.}
at 430~MHz.  
Spectra across a 
10~MHz band centered on this frequency were produced by a 
Fourier-transform-based spectrometer 
(AOFTM\footnote{http://www.naic.edu/\~\,aoftm}).
RCP and LCP signals were 
summed in hardware, and
 spectra with 1024 
frequency channels (i.e. 10~kHz frequency resolution)
 were produced every $\approx$~1.6~ms.
During later analysis the spectra were 
synchronously averaged at the pulsar period, calculated using TEMPO \cite{tay89},
and summed over 10~s time intervals,
creating a cube 
with axes of time, frequency, and pulsar rotational phase. We used 512
phase bins across the pulsar period.  Dynamic spectra
were created by producing an ON-pulse spectrum from a window of 
phase bins centered on the pulse and subtracting an OFF-pulse spectrum  
to remove the baseline.

We observed 19 pulsars 
during 1999 January and 11 pulsars during 2000 January.
A typical observation lasted one hour; we made
46 and 43 separate observations at the two observing epochs, respectively.
Following several previous studies \nocite{cw86} (\eg Cordes \& Wolszczan 
1986; RLG) we produced two-dimensional 
power spectra from the dynamic spectra.
These ``secondary spectra'' form the basis for most of the further 
analysis reported here.   
We normalized each secondary spectrum by its 
maximum and converted the relative power levels into a 
decibel scale.

Figure~\ref{fig:dynsec} shows dynamic and secondary spectra for
PSR~B0834+06. The crisscross pattern in
the dynamic spectrum shows up in the secondary spectrum as two 
parabolic features, curving away from the conjugate time axis,
that we call ``arcs''.
These arcs are remarkable for their 
narrowness and extent. In Figure~\ref{fig:coadded} we present secondary spectra,
averaged over multiple days to improve dynamic range.
PSR~B0823+26 and PSR~B0834+06 showed a persistent 
arc pattern over both the 1999 and 2000 observing runs.
In 2000, PSR~B0919+06 showed arcs that were relatively broad and 
faint and, hence,  showed up only upon averaging of $\sim 6$ secondary spectra.
However, we did not see arcs in
two high-quality 1999 observations. 
PSR~B1133+16 showed arcs during most of the 1999 and 2000
observations, but several high-quality 2000 observations 
failed to show them, even though they were made on days adjacent 
to those on which they were present.
PSR~B1933+16 did not show the arc pattern in either of 
two high-quality observations in 1999 or 2000.
Time variability of the arcs 
will be the subject of a separate study.

\section{Results}
We detected arcs in the secondary spectra of pulsars
B0823+26, B0834+06, B0919+06, and B1133+16.
By using the relationship between fringe frequencies in the secondary 
spectrum and scattering parameters 
(\eg Cordes \& Wolszczan 1986; RLG; Gupta, Bhat, \& Rao 1999),  
we can analyze the arcs more 
quantitatively.
Here we summarize a more detailed study that will be reported elsewhere
(Cordes \etal 2001).
\nocite{ccrs00}
To simplify calculations, these treatments assume
that the scattering takes place in a thin screen a 
distance $D_{s}$ from the pulsar, with the pulsar-observer distance 
$D$.  They show that the interference between two points  $\thvec_{1,2}$ on the image 
plane (i.e. as seen by an observer) will 
lead to two-dimensional fringing in the dynamic spectrum with fringe 
frequencies
\be
f_t &=& 
  \left(\frac{D}{\lambda\ds}\right)(\thvec_2 - \thvec_1)\cdot\veffperpvec,
	\label{eq:ft}		\\
f_{\nu} &=&  \left( \frac{D}{2c\ds}\right)
		(D-\ds) \left(\thvec_2^2 - \thvec_1^2\right).
	\label{eq:fnu}
\ee
Here $f_{t}$ is the conjugate time axis, $f_{\nu}$ is 
the conjugate frequency axis, and the effective velocity (Cordes \& Rickett 1998)
\nocite{cr98} 
is a weighted sum of the velocities
of the source, screen and observer, 
\be
\veffperpvec = (1 - \ds/D)\vpperpvec + (\ds/D)\vobsperpvec  
     -\vismperpvec .
\ee
We can introduce the dimensionless variables
\be
p &=& \left(\thvec_2^2 - \thvec_1^2\right)  
	=  \left [\frac{2c\ds}{D(D-\ds)}\right ] f_{\nu}, 
	\label{eq:pbar}\\
q &=& (\thvec_2 - \thvec_1)\cdot\veffperpvechat 
	= \left[\frac{\lambda\ds}{D\veffperp}\right] f_t, 
	\label{eq:qbar}
\ee
where $\veffperpvechat$ is a two-dimensional unit vector for the effective
velocity.   
Since $p$ is quadratic in image 
angles and $q$ is linear in image angles, $p$ will generally be 
quadratic in $q$.  This is essential to explaining the 
parabolic curvature of the arcs. 

If we make the reasonable assumption that the pulsar velocity is much 
greater than either the velocity of the Earth or the screen, 
equations~(\ref{eq:pbar})--(\ref{eq:qbar})
can be simplified to 
\be
p &=& \left [\frac{2 c s}{D(1-s)}\right ] f_{\nu}, 
	\label{eq:pnew}\\
q &=& \left[\frac{\lambda s}{V_{\rm pm} (1-s)}\right] f_t, 
	\label{eq:qnew}
\ee
where $s \equiv \ds/D$, and we approximate  $\veffperp$ by the measured proper motion
speed,  $V_{\rm pm} = \mu D$.
A parabolic arc ($p=q^{2}$) in the p-q plane then corresponds to a
feature in the secondary spectrum of curvature
\be
f_{\nu} &=& \frac{D}{2c} \frac{\lambda^{2}}{V_{\rm pm}^{2}}
           \left(\frac{s}{1-s}\right)  f_{t}^{2}.
           \label{eq:dyn-arc}
\ee
In this model, the curvature of the arcs is determined
solely by the screen placement,
the pulsar distance, and the pulsar proper motion.
Thus, by measuring the curvature of the arcs in the secondary spectrum,
we can determine a value for the screen placement that is 
consistent with the observations and assumed values of $D$ and $V_{\rm pm}$,
as in Table~\ref{tab:screen}.

Gupta (1995)
\nocite{gup95}
used
diffractive
bandwidth and timescale measurements to determine
$V_{\rm iss}$, the 
speed of the scintillation pattern past the observer (Cordes and 
Rickett 1998).
\nocite{cr98}
Assuming that the scattering is dominated by a thin screen, 
one can determine the 
screen placement for which 
$V_{\rm iss}$ is consistent with
$V_{\rm pm}$.
This is determined from Gupta's $R_{1}$ 
parameter by $s_{\rm iss}=R_{1}^{2}/(1+R_{1}^{2})$.
Table~\ref{tab:screen} shows that the values of $s_{\rm iss}$ agree 
well with those for $s_{\rm arc}$, calculated with Eq. 8.
Although the values of $s_{\rm iss}$ agree well with
those for $s_{\rm arc}$, the agreement may be only proportionate because
there are significant uncertainties in both the distances and the ISS
parameters used by Gupta; also, scattering is likely to receive contributions
from the distributed ISM as well as from a screen.
In particular, a recent study of B0919+06 (Chatterjee \etal 2001) that
compares
scintillation velocity estimates, which are epoch dependent, and
the parallax and proper motion determined from VLBA observations,
finds that a thin screen must be combined with a distributed medium such as
that
of the Taylor \& Cordes (1993) model to bring agreement between the two
velocity estimates.   At some epochs, the screen evidently causes multiple
imaging of the pulsar. 

From equation~(1) it can be seen that 
the vertical ($f_{t}$) extent of arc features in the 
secondary spectrum is related to the angular extent of the scattering 
image along the effective velocity vector.  
Table~1 shows the typical angular extent of the 
filament, $\theta_{x{\rm ,typ}}$, and the corresponding linear size on 
the screen, $x_{\rm typ}$, inferred from the value of $f_{t}$ for a 
generic model.

We analyze the distribution of power in 
the secondary spectrum with  a sequence of vertical cuts, as in
Figure~\ref{fig:cuts}. 
The rapid decline of power above the arcs strongly limits
the image geometries that are consistent with the data.
In order to exclude power above the arc,
points must satisfy $p \ge q^{2}$.
Referring to equations~(\ref{eq:pbar})--(\ref{eq:qbar}), this 
constraint can be expressed as
\be 
\theta_{2}^{2}-\theta_{1}^{2} &\ge& (\theta_{2{\rm x}}-\theta_{1{\rm x}})^{2},
             \label{eq:inequality}
\ee
where the $x$~components are 
along the effective velocity vector.
A single extended image has many image pairs that violate this 
condition and, hence, will produce a 
large amount of power above the arcs, 
inconsistent with the behavior seen.
Hence, a multiple image geometry is required to produce the clearly 
delineated arcs we observe.

One configuration that satisfies Eq. 9, having no detectable power above 
the arcs, consists of a bright 
point-like image at the origin (direct line of sight to the pulsar)
and a faint secondary image as a filled halo around it.
The interference between a point at the origin and any other 
image point cannot produce power above the arcs 
because $\thvec_{1} = 0$.
If the point-like image and halo have flux densities 
proportional to $a$ and $b$, 
respectively, 
then the point-like source will produce a point-like feature at the 
origin of the secondary spectrum with power proportional to $a^{2}$ 
(\eg Cordes \& Wolszczan 1986). 
The inter-image interference between the point-source and the halo will 
produce power proportional to $a b$.  The 
self-interference between regions of the halo (which {\it will} produce 
power above the arcs) will be proportional to 
$b^{2}$.  Since this is smaller by another factor of $b/a$ from 
the inter-image term, then, for small $b/a$, the only detectable
power will be a point-source at the 
origin and power below the arcs.

Although the point-source plus halo model is consistent with the
observations, the scattering 
material in the screen would need to be inhomogeneous enough that the 
direct image of the pulsar would shine through, but rays slightly 
off-axis would be scattered.  In essence, we would need to be seeing 
the pulsar through a break in the interstellar ``clouds'',  consistent with the
time-variability of the arcs.
On the other hand, there is strong evidence for 
scattering geometries based on 
filamentary structures \cite{ric90}. However, the self-interference between
filaments will lead to features which are inconsistent with our data.
Any viable model based on 
filamentary features in the image plane must somehow hide these 
self-interference features as well as produce the observed parabolic 
arcs.

The formalism used here includes only geometric delays, 
whereas 
differential dispersive delays may
be important in some cases.
The fact that the
observed arcs conform to the geometric-only formalism implies that 
at least in some instances
the dispersive effects are small.
We have examples of secondary spectra that show more complex behavior
and will analyze them within the more general framework in a later paper.
An extended medium, as opposed to a single screen, also will be explored.

While there are many possibilities for source geometries, it is clear 
that a multicomponent scattered image is necessary.  
In future 
papers, we will apply 
this method to better understand the 
structure of the ISM on small spatial scales.

We thank T.~Joseph~Lazio for helpful discussions, the operators at Arecibo for
help with remote observations, and Duncan Lorimer for the development of an excellent
remote observing interface.
We acknowledge support from the National Science Foundation:  
grant AST-9618408 to JMC and grant
AST-9619493 to DRS.


\begin{thebibliography}{}

\bibitem[Chatterjee \etal (2001)]{ccl+00}
Chatterjee, S., Cordes, J.~M., Lazio, T.~J.~W., Goss, W.~M., Fomalont, E.~B.
\& Benson, J.~M. 2001, \apj, submitted

\bibitem[Cordes \etal (2001)]{ccrs00}
Cordes, J.~M. \etal 2001, in preparation

\bibitem[Cordes \& Rickett (1998)]{cr98}
Cordes, J.~M. \& Rickett, B.~J., 1998, \apj, 507, 846

\bibitem[Cordes \& Wolszczan (1986)]{cw86}
Cordes, J.~M. \& Wolszczan, A., 1986, \apjlett, 307, L27

\bibitem[Cordes, Weisberg \& Boriakoff (1985)]{cwb85}
Cordes, J.~M., Weisberg, J.~M.  \& Boriakoff, V., 1985, \apj, 288, 221

\bibitem[Gupta (1995)]{gup95}
Gupta, Y., 1995, \apj, 451, 717

\bibitem[Gupta, Bhat, \& Rao (1999)]{gbr99}
Gupta, Y., Bhat, N.~D.~R. \& Rao, A.~P. 1999, \apj, 520, 173

\bibitem[Gupta, Rickett \& Lyne (1994)]{grl94}
Gupta, Y., Rickett, B.~J.  \& Lyne, A.~G., 1994, \mnras, 269, 1035

\bibitem[Harrison, Lyne \& Anderson (1993)]{hla93}
Harrison, P.~A., Lyne, A.~G.  \& Anderson, B., 1993, \mnras, 261, 113

\bibitem[Hewish (1980)]{hew80}
Hewish, A., 1980, \mnras, 192, 799

\bibitem[Hewish, Wolszczan \& Graham (1985)]{hwg85}
Hewish, A., Wolszczan, A.  \& Graham, D.~A., 1985, \mnras, 213, 167

\bibitem[Rickett (1990)]{ric90}
Rickett, B.~J., 1990, \araa, 28, 561

\bibitem[Rickett, Lyne \& Gupta (1997)]{rlg97}
Rickett, B.~J., Lyne, A.~G.  \& Gupta, Y., 1997, \mnras, 287, 739 
(RLG)

\bibitem[Roberts \& Ables (1982)]{ra82}
Roberts, J.~A. \& Ables, J.~G., 1982, \mnras, 201, 1119

\bibitem[Scheuer (1968)]{sch68}
Scheuer, P. A.~G., 1968, Nature, 218, 920

\bibitem[Taylor \& Weisberg (1989)]{tay89} Taylor, J. H. \& Weisberg, J. M. 1989,
\apj, 345, 132

\bibitem[Taylor \& Cordes (1993)]{tc93}
Taylor, J.~H. \& Cordes, J.~M., 1993, \apj, 411, 674

\bibitem[Wolszczan \& Cordes (1987)]{wc87}
Wolszczan, A. \& Cordes, J.~M., 1987, \apjlett, 320, L35

\end{thebibliography}


\begin{deluxetable}{lcccccccc} 

\tablecolumns{9} 

\tablewidth{0pc} 

\tablecaption{Scattering Screen Location \label{tab:screen}} 

\tablehead{ 
\colhead{PSR} & \colhead{Distance}   & \colhead{$V_{\rm pm}$} &
    \colhead{$f_{\nu{\rm ,arc}}$} & \colhead{$f_{t {\rm ,arc}}$} &
    \colhead{$s_{\rm arc}$} &
    \colhead{$s_{\rm iss}$} & \colhead{$\theta_{x{\rm ,typ}}$}  &
    \colhead{$x_{\rm typ}$} \\
\colhead{} & \colhead{(kpc)}   & \colhead{(km/s)} &
    \colhead{(${\rm MHz^{-1}}$)} & \colhead{(${\rm min^{-1}}$)} &
    \colhead{}   & \colhead{} & \colhead{(mas)}  & \colhead{(AU)}}
\startdata 

B0823+26 & 0.38 & 196 & 26 & 0.8 & 0.36 & 0.40 & 5.4 & 1.3\\

B0834+06 & 0.72 & 174 & 31 & 0.6 & 0.33 & 0.36 & 4.0 & 1.9\\

B0919+06 & 1.2 & 505 & 15 & 0.55 & 0.59 & 0.63 & 3.6 & 1.8\\

B1133+16 & 0.27 & 475 & 11 & 1.15 & 0.49 & 0.46 & 5.4 & 0.7\\
\enddata 

\tablecomments{Values are derived from the Taylor and Cordes 
(1993) distance model and the Harrison \etal (1993) proper
motions, except for 
recent interferometric results for PSR~B0919+06
(Chatterjee \etal 2001).
\nocite{tc93}
\nocite{ccl+00}
\nocite{hla93}
}
\end{deluxetable} 


\figcaption[Figure1.ps]{A dynamic spectrum (top) and its secondary 
spectrum for a 30 min observation of PSR~B0834+06 on 2000~January~18.  
The flux density 
is displayed as a linear gray-scale with black indicating highest flux.
The Nyquist limit along the vertical axis of the secondary spectrum is 3~cycles/min, but the 
spectrum is noise-like beyond the region shown.  The gray-scale is 
logarithmic (linear in decibels),
with the white level set 3~dB above the noise and the 
black level set 5~dB below the maximum array value at the origin,
as in RLG.  These limits
span a 48~dB range.
The strong vertical line in the center of the plot is due partly
to the sidelobe response of the Fourier transform and partly to
broadband pulse-to-pulse fluctuations that are only moderately
quenched by the 10-s averaging interval.
A similar horizontal line is due primarily to the sidelobe response.
\label{fig:dynsec}}  

\figcaption[Figure2.ps]{Co-added secondary spectra for 5 pulsars.  The 
logarithmic gray-scale is set as in Figure~\ref{fig:dynsec}.  
Only a 
portion of the conjugate time axis is shown in several cases.
The range of displayed values is 49,
52, 48, 44, 56, and 39 dB for panels (a)-(f), respectively.
Panels (c) and (d) illustrate the time variability of 
the arcs; panel (f) is an example 
of a strong pulsar that does not display arcs.
\label{fig:coadded}}
   
\figcaption[Figure3.ps]{Cross-sectional cuts through the secondary spectrum
of Figure~\ref{fig:coadded}(b), which was first smoothed by
1~cycle/MHz and 0.06~cycles/min ($10 \times 10$ pixels).
As shown by the tickmarks in Figure~\ref{fig:coadded}(b), the cuts are taken 
parallel to the vertical axis at steps of 2~cycles/MHz.
The cuts are made by reflecting the 
second quadrant of the secondary spectrum through the origin into the 
fourth quadrant and then cutting vertically along 
the conjugate time axis.  The region near where the cuts cross 
$f_{t}=0$ has been set to a constant value since the sidelobes 
rise to about $-10$~dB there.  Only the first six 
cuts are labelled.
Salient features are the similar 
height of the arcs 
on both sides of the origin, the sharp drop-off of power above 
the arcs (i.e. $|{f_t}| \ge 0.5$; note the logarithmic axis), the nearly constant power level below the 
arcs, and the low value of the noise floor.
\label{fig:cuts}}

\end{document}